# A Complex-Network Perspective on Alexander's Wholeness

Bin Jiang

Faculty of Engineering and Sustainable Development, Division of GIScience
University of Gävle, SE-801 76 Gävle, Sweden
Email: bin.jiang@hig.se

*(Draft: February 2016, Revision: April, July 2016)*

> *"Nature, of course, has its own geometry. But this is not Euclid's or Descartes' geometry. Rather, this geometry follows the rules, constraints, and contingent conditions that are, inevitably, encountered in the real world."*
>
> Christopher Alexander et al. (2012)

**Abstract**
The wholeness, conceived and developed by Christopher Alexander, is what exists to some degree or other in space and matter, and can be described by precise mathematical language. However, it remains somehow mysterious and elusive, and therefore hard to grasp. This paper develops a complex network perspective on the wholeness to better understand the nature of order or beauty for sustainable design. I bring together a set of complexity-science subjects such as complex networks, fractal geometry, and in particular underlying scaling hierarchy derived by head/tail breaks – a classification scheme and a visualization tool for data with a heavy-tailed distribution, in order to make Alexander's profound thoughts more accessible to design practitioners and complexity-science researchers. Through several case studies (some of which Alexander studied), I demonstrate that the complex-network perspective helps reduce the mystery of wholeness and brings new insights to Alexander's thoughts on the concept of wholeness or objective beauty that exists in fine and deep structure. The complex-network perspective enables us to see things in their wholeness, and to better understand how the kind of structural beauty emerges from local actions guided by the 15 fundamental properties, and by differentiation and adaptation processes. The wholeness goes beyond current complex network theory towards design or creation of living structures.

## 1. Introduction

Nature, or the real world, is governed by immense orderliness. The order in nature is essentially the same as that in what we build or make, and underlying order-creating processes of building or making of architecture and design are no less important than those of physics and biology. This is probably the single major statement made by Alexander (2002–2005) in his theory of centers, in which he addressed the fundamental phenomenon of order, the processes of creating order, and even a new cosmology – a new conception of how the physical universe is put together. In the theory of centers or living geometry (Alexander et al. 2012), the wholeness captures the meaning of order and is defined as a life-giving or living structure that appears to some degree in every part of space and matter; see Section 3 for an introduction to wholeness and wholeness-related terms. As the building blocks of wholeness, centers are identifiable coherent entities or sets that overlap and nest each other within a larger whole. Unlike the previous conception of wholeness focusing on the gestalt of things (Köhler 1947), the wholeness of Alexander (2002-2005) is not just about cognition and psychology, but something that exists in space and matter. Different from the wholeness in quantum physics mainly for understanding (Bohm 1980), Alexander's wholeness aims not only to understand the phenomenon of order, but also to create order in the built world or art. The wholeness is defined as a recursive structure. Based on this definition, Jiang (2015) developed a mathematical model of wholeness as a hierarchical graph with indices for measuring degrees of life or beauty for both individual centers and the whole. This model helps address not only why a design is beautiful, but also how much beauty it has. However, this previous study had some fundamental issues on the notions of centers and wholeness unaddressed. Specifically, what are the centers, how are they created, and how do they work together to contribute to the life of wholeness? In addition, the wholeness remains somehow mysterious, particularly within our current mechanistic worldview (Alexander 2002–2005). To address these fundamental issues, this paper develops a complex-network perspective on the wholeness.

A complex network is a graph consisting of numerous nodes and links, with unique structures that differentiate it from its simple counterparts such as regular and random networks (Newman 2010). Simple networks have a simple structure. In a regular network, all nodes have a uniform degree of connectivity. In a random network, the degrees of connectivity only vary slightly from one node to another. As a consequence, a random network hardly contains any clusters, not to say overlapping or nested clusters. On the contrary, complex networks, such as small-world and scale-free networks (Watts and Strogatz 1998, Barabási and Albert 1999), tend to contain many overlapping and nested clusters that constitute a scaling hierarchy (Jiang and Ma 2015; see a working example in Section 2). The scaling hierarchy is a distinguishing feature of complex networks or complex systems in general. For example, a city is a complex system, and a set of cities is a complex system (Jacobs 1961, Alexander 1965, Salingaros 1998, Jiang 2015c), both having scaling hierarchy seen in many other biological, social, informational, and technological systems. This paper demonstrates that the wholeness bears the same scaling hierarchy as complex networks or complex systems in general.

Relying on the complex network perspective, this paper aims to demonstrate that wholeness is not just in cognition and psychology, but something that exists in space and matter. It also aims to show that the concept of wholeness is important not just for understanding the phenomenon of order, but also for creating order with a high degree of wholeness through two major structure-preserving transformations: differentiation and adaptation. I argue that there are major differences between the whole as a vague term and wholeness as a recursive structure. The wholeness comprises recursively defined centers induced by itself, whereas the whole, as we commonly perceive, comprises pre-existing parts. The mantra that *the whole is more than the sum of its parts* should be more truly rephrased as *the wholeness is more than the sum of its centers*. This paper examines the notions of wholeness and centers from the perspective of complexity science. It also discusses two types of coherence respectively created by differentiation and adaptation processes, which are consistent with the spatial properties of heterogeneity and dependence for understanding the nature of geographic space.

The remainder of this paper is structured as follows. Section 2 briefly introduces complex networks and scaling hierarchy using head/tail breaks – a classification scheme and a visualization tool for data with a heavy-tailed distribution (Jiang 2013a, Jiang 2015a). Section 3 compares related concepts, such as whole and parts versus wholeness and centers, and discusses the theory of centers using two examples of a cow and an IKEA desk. Section 4 presents three case studies to show how wholeness emerges from space and how it can be generated through the two major structure-preserving transformations of differentiation and adaptation. Section 5 further discusses implications of the complex-network perspective and wholeness. Finally, Section 6 draws a conclusion and points to future work.

## 2. Complex networks and the underlying scaling hierarchy
Small-world and scale-free networks are two typical examples of complex networks, which fundamentally differ from their regular and random counterparts. A small-world network is a middle status between the regular and random networks, so it has some nice properties of its regular and random counterparts. These properties are local efficiency of regular networks characterized by high clustering coefficient, and global efficiency of the random networks measured by short average path length (Watts and Strogatz 1998). Scale-free networks are a special type of complex networks, and their degree of connectivity demonstrates a power-law distribution, indicating far more less-connected nodes than well-connected ones (Barabási and Albert 1999). In other words, very few nodes, or hubs, have the highest degree of connectivity, many nodes have the lowest degree, and some in between the highest and the lowest. Complex networks are so called precisely because of their complex structure that involves a large number of nodes and components. The nodes are the basic units, while the components are those built from the basic units.

Complex networks tend to contain many components, termed communities or clusters (Newman 2004). A cluster has many inside links and a few outside links, so constitutes a coherent sub-whole or sub-structure that adapts to its context. Clusters within a complex network nest each other, forming a scaling hierarchy of far more small communities than large ones (Jiang and Ma 2015). To illustrate, we adopt the Karate Club network (Zachary 1977), widely studied in the literature of social and complex networks. This network, consisting of 34 nodes and 78 links, can be broken down into 14 communities of different sizes: 28, 15, 10, 6, 5, 5, 3, 2, 2, 2, 1, 1, 1 and 1 (Figure 1). Apparently there are far more small communities than large ones. There are also many nested relationships. For example, the community of size 28 consists of a community of size 15, which is further broken down into communities of sizes 6, 5, 2, 1, and 1. The scaling hierarchy is formed from the clusters, which are similar to human organs that function as independent coherent parts, but still fit into the context of the human body. More generally, a complex network is not assembled from mechanical parts, but is more like a tree or human body growing or unfolding from a seed or an embryo. From the design point of view, clusters as coherent parts are integrated into their contexts as a whole (Alexander 1964). This introduces the notion of adaptation or fit, which recurs between adjacent elements and systems, and helps create harmony or coherence at individual local scales. Clusters resemble the concept of centers (see Section 3 for more details) as a building block of the wholeness.

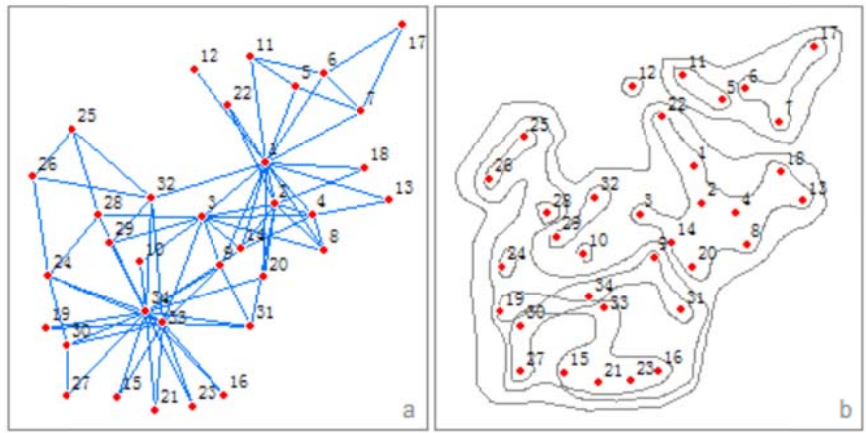

Figure 1: (Color online) The Karate Club network broken into mutually nested communities
(Note: Panel a shows the social network, consisting of 34 nodes and 78 links. Panel b depicts nested relationships of the 14 communities of sizes 28, 15, 10, 6, 5, 5, 3, 2, 2, 2, 1, 1, 1 and 1 with far more small communities than large ones. These communities were detected by an algorithm inspired by head/tail breaks (Jiang 2013a, Jiang 2015a).)

The underlying scaling hierarchy is an important property of complex networks or complex systems in general (Simon 1962). It appears in a variety of natural and societal complex systems such as social, biological, technological, and informational ones. This scaling hierarchy is not tree-like but rather a complex network with many redundant and overlapping links. This insight was originally observed through naturally evolved cities, as Alexander (1965) nicely articulated in the classic *A City is Not a Tree*. He used the term semi-lattice to refer to the scaling hierarchy simply because the concept of complex networks was not known then. In this respect, Alexander was far ahead of the time when complex networks were understood. More importantly, his thoughts are not just limited to understanding surrounding things, just as what current complex network theory does, but aim for creating things with living structure (i.e., buildings, communities, cities or artifacts). The scaling hierarchy can be further seen from the relevance or importance of individual nodes within a complex network. Through Google's PageRank (Page and Brin 1998), nodes' relevance or importance can be computed, as there are far more less-important nodes than more-important ones. PageRank is recursively defined and resembles the wholeness as a recursive structure. This will be further discussed in the following section.

**3. The wholeness and the theory of centers**
The general idea of wholeness, or seeing things holistically, can be traced back to the Chinese philosopher Chuang Tzu (360 BC), who saw the structure of a cow as a complex whole, in which some parts were more connected (or coherent) than others. The butcher who understands the cow's structure always cuts the meat from the soft spots and the crevices of the meat, so makes the meat fall apart according to its own structure. The butcher therefore can keep his knife sharp for a hundred years. In the 20$^{th}$ century, wholeness was extensively discussed by many writers prominent in Gestalt psychology (Köhler 1947), quantum physics (Bohm 1980), and many other sciences such as biology, neurophysiology, medicine, cosmology and ecology. However, none of these writers prior to Alexander (2002–2005) showed how to represent or formulate wholeness in precise mathematical language. According to the Merriam-Webster dictionary, whole is something complete, without any missing parts. On the other hand, wholeness is the quality of something considered as a whole. However, both whole and wholeness have deeper meanings and implications in the theory of centers (Alexander 2002–2005). In particular, the notion of wholeness is so subtle and profound that Alexander (1979) previously referred to it as 'the quality without a name'. He struggled with different names, such as alive, comfortable, exact, egoless and eternal, but none of these captured the true meaning of the quality. In this paper, as in Alexander (2002-2005), the three terms wholeness, life, and beauty are interchangeably used when appropriate to indicate order or coherence. Things with a high degree of wholeness are called living structure.

Table 1: Comparison of whole and wholeness using two examples of a cow and an IKEA desk

| **Whole (noun + adjective)** | **Wholeness (structure + measure)** |
|---|---|
| on the surface | deeper below the surface |
| consists of parts *(see below subsection)* | consists of centers *(see below subsection)* |
| easy to see, the whole of a cow is the cow itself | hard to see as a recursive structure, the wholeness of a cow |
| hard to sense, one thing is more whole than another | hard to sense as the degree of coherence somehow like temperature |
| easy to sense, a cow is more whole than a desk | easy to sense, a cow has a higher wholeness than a desk |
| a whole assembled from parts like a desk | a desk has low wholeness or low coherence |
| a whole unfolded from an embryo like a cow | a cow has high wholeness or high coherence |
| *Parts* | *Centers* |
| pre-existing in the whole | induced by the wholeness |
| on the surface | deeper below the surface |
| easy to see | hard to see |
| mechanical | organic |
| non-recursive | recursive |
| simple | complex |

Wholeness is defined as a recursive life-giving structure that exists in space and matter, and it can be described by precise mathematical language (Alexander 2002-2005). Although whole and wholeness seem different, they are closely related, and sometimes refer to the same thing. The whole is on the surface and is referred to informally, while wholeness is below the surface and is referred to formally (Table 1). The term whole can be both a noun and an adjective. For example, a cow is a whole, and a cow is more whole than a desk. On the other hand, wholeness has two different meanings: a recursive structure, and as a measure for degree of wholeness or coherence. A whole is easy to see – being a relatively coherent spatial set, because it appears on the surface. For example, the whole of a cow is the cow itself. The wholeness as a recursive structure is hard to see, since it lies deeper consisting of atoms, molecules, cells, tissues, and organs forming a scaling hierarchy. It is usually difficult to sense that one thing has a higher degree of wholeness, or is more whole than another. However, things assembled from parts are often less whole than things grown from embryos. For example, a cow has a higher degree of wholeness than a desk. We can also compare the two related terms parts and centers, of which the whole and wholeness respectively consist. Parts usually pre-exist in the whole and refer to mechanical pieces that are easy to see because they appear on the surface and are non-recursive and simple. Centers are mainly created by the wholeness and refer to organic pieces that are hard to see because they exist deeper below the surface, and are recursive and complex. The differences between the parts and centers show two different world views: the mechanistic, to which we are accustomed, and the organic, which underlies Alexander's radical thought.

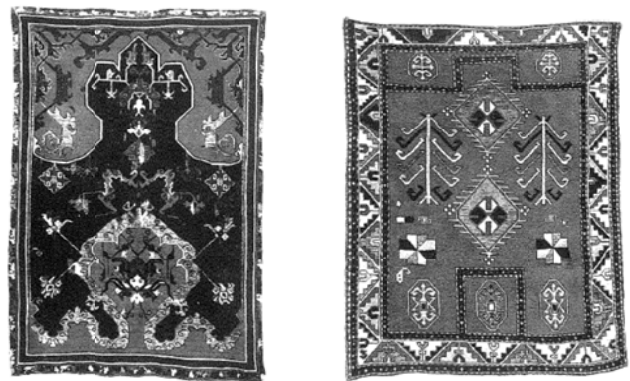

Figure 2: Two carpets with different degree of wholeness (Alexander 1993)
(Note: The left is more whole than the right, or the left has a higher degree of wholeness than the right.)

Table 2: The 15 fundamental properties of wholeness

| | | |
|---|---|---|
| Levels of scale | Good shape | Roughness |
| Strong centers | Local symmetries | Echoes |
| Thick boundaries | Deep interlock and ambiguity | The void |
| Alternating repetition | Contrast | Simplicity and inner calm |
| Positive space | Gradients | Not separateness |

To further elaborate on the wholeness, let us examine which of the two carpets (Figure 2) is more whole, or has a higher degree of wholeness. Both the carpets possess a high degree of wholeness, but the left one has a higher degree, or is more whole, than the one on the right. This phenomenon of wholeness can be captured through the mirror-of-the-self experiment (Alexander 2002-2005). Two objects or their images are put side-by-side in front, and you are asked to choose one that represents a picture of your own deepest or truest self as a whole. The experiment is not to choose one that you prefer, which is likely to be idiosyncratic, accounting for 10 percent of our feelings. Instead, you must pick one that reflects your own inner self. This part of human experience, accounting for 90 percent of our feelings, is shared among people and across cultures (Alexander 2002–2005). The wholeness exists both physically in the world and psychologically inside the human self. Physically, the phenomenon of wholeness comprises the 15 fundamental properties such as levels of scale, strong centers, boundaries, and local symmetries (see Table 2). The left carpet has more of the 15 properties than the right, or more of the mirror-of-the-self qualities than the right. The mirror-of-the-self experiment is somehow like relying on human sense to compare two temperatures. Our feelings cannot provide a precise measurement of temperature. Equally, it is usually hard to compare the wholeness of two complex things such as a cow and a tree. In this regard, the mathematical model of wholeness (Jiang 2015b), relying on a hierarchical graph for representing the wholeness, can accurately measure the degree of wholeness or life. The model is essentially a complex-network perspective on wholeness.

**4. The wholeness from a complex-network perspective**
This section presents three case studies to show how wholeness is represented as a complex network of its centers, and how centers are created by wholeness. We begin with the simplest case study of a paper with a tiny dot, followed by studies of the Alhambra plan and the Sierpinski carpet. Through the case studies, we demonstrate that wholeness exists in space. More importantly, we elaborate on how whole or wholeness can be created step by step by following the 15 fundamental properties or structure-preserving transformations.

**4.1 A paper with a tiny dot**
The wholeness is very subtle, but also very concrete. To discuss wholeness, Alexander (2002–2005) presented a simple example involving a blank paper with a tiny dot (Panels a and b of Figure 3). I use the same example to examine how the wholeness emerges and changes before and after a tiny dot is placed on the paper. A blank paper is a whole, which is easy to see. The wholeness of the blank paper is not hard to imagine because it is pretty simple, consisting of four corners and the gravitational center of the paper. However, after a tiny dot (no more than 0.0001 of the sheet) is placed, the whole remains without much change on the surface, but its wholeness changes dramatically. There are at least 20 latent centers, such as a halo around the dot, the four rectangles around the dot, the four corners, and different rays from the dot to the corners (Panels c-h of Figure 3). Among the centers, only the paper and dot pre-exist. All others are induced by the overall configuration of space or wholeness. Seen in Figure 3, these centers are real, not just in cognition and psychology. There are also supporting relationships among the centers, which constitute a complex network as a whole (Figure 4a). The whole or the overall configuration of the space is the source of the centers' strength, as indicated by the dot sizes.

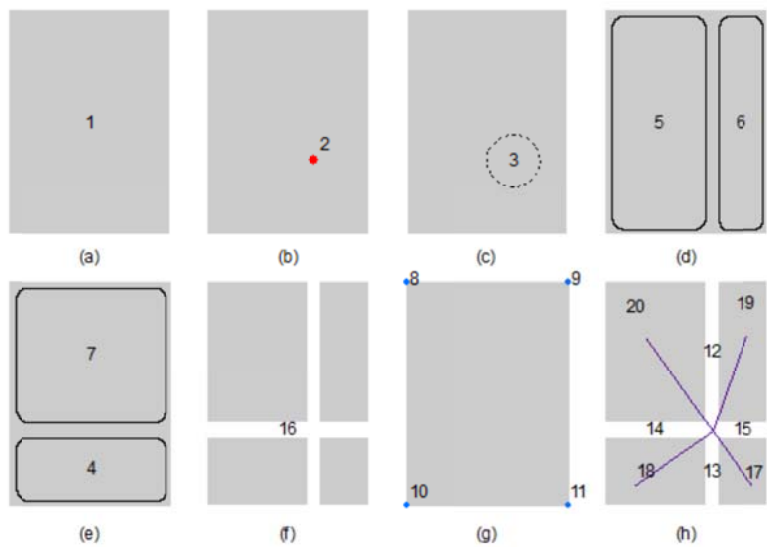

Figure 3: (Color online) A simple wholeness emerged from a blank paper with a dot placed
(Note: The dot in the sheet of paper created at least 20 entities, listed according to their relative strength: (1) the paper itself, (2) the dot, (3) the halo, (4) the bottom rectangle, (5) the left-hand rectangle, (6) the right-hand rectangle, (7) the top rectangle, (8) the top left corner, (9) the top right corner, (10) the bottom left corner, (11) the bottom right corner, (12) the ray going up, (13) the ray going down, (14) the ray going left, (15) the ray going right, (16) the white cross by these four rays, (17) the diagonal ray to the bottom right corner, (18) the diagonal ray to the bottom left corner, (19) the diagonal ray to the top right corner, and (20) the diagonal ray to the top left corner.)

The paper with the dot is more whole than the blank paper itself. The wholeness of the original blank paper constitutes a simple network containing only five nodes (Figure 4b). However, the network representing the wholeness of the paper with the dot is far more complex or whole than that of the blank paper (Figure 4a). First, the whole of the blank paper contains only five nodes, whereas the whole of the paper with the dot contain the 20 nodes. Second, the 20 nodes form a striking scaling hierarchy, i.e., far more less-connected nodes than well-connected ones. Third, the 20 nodes constitute different communities, such as the four corners, the four rays, and the top three nodes. Seen in Figure 4a, the strongest center number 1 is followed by centers 2, 3, and so on. There are several sub-wholes, such as centers 8, 9, 10, and 11 as corners; 12, 13, 14, and 15 as rays; and 17, 18, 19, and 20 as diagonal rays. These sub-wholes constitute some implicit relationships within the complex network.

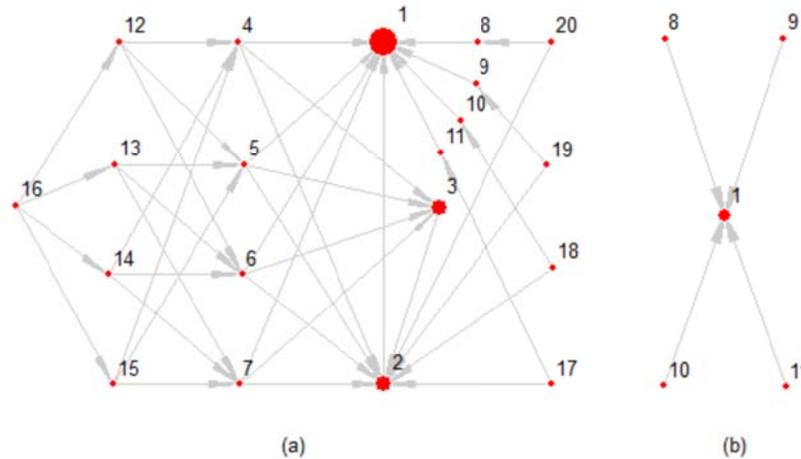

Figure 4: (Color online) The networks of the wholeness respectively emerged from (a) the paper with the dot, and (b) the blank paper
(Note: The first network consists of the 20 nodes, while the second network consists of only the four nodes, both with dot sizes indicating their strength. It is obvious that the paper with the dot is more whole, or has a higher degree of wholeness, than the blank paper.)

Compared to the complex network presented in Section 2, the networks for the wholeness are directed. As a rule, weak centers tend to support strong ones, as shown in Figure 4. This kind of support relationship is cumulative or iterative. This is why center 1 is the strongest, with nine in-links, which are further enhanced by some in-links in an iterative manner. Although center 2 has also nine in-links, it is weaker than center 1 because of the iterative nature of support relationships. The centers of 16, 17, 18, 19 and 20 are the weakest, because they only contain out-links. The hierarchy shown in Figure 4a is very steep, while the one in Figure 4b is very flat.

**4.2 The Alhambra plan**
The Alhambra plan possesses many of the 15 fundamental properties that help identify many latent centers. Thick boundaries, local symmetries, levels of scale, and strong centers are probably the most salient properties. There are at least 720 positive spaces or centers, and 880 relationships (Jiang 2015b, Figure 5). There are more other centers created by the wholeness or the overall configuration of the space. For example, the whole of the plan consists of many nested sub-wholes or centers defined at many different levels. It consists of three sub-wholes at the first level, which can be named as the left sub-whole, the middle sub-whole, and the right sub-whole. Each of the three sub-wholes is asymmetrical, containing further three sub-wholes, so there are nine sub-wholes at the second level. The left sub-whole comprises three sub-wholes: left, middle and right. Each of the middle and right sub-wholes comprises three sub-wholes: top, middle and bottom. The nine sub-wholes at the second level can be further broken down into sub-wholes or centers at the third or fourth level, recursively. The whole and its numerous sub-wholes (or centers) constitute mutually reinforcing and supporting relationships, usually with small surrounding centers pointing to big central ones. Scaling hierarchy is eventually formed, and it is the source of life or beauty of the building complex.

The plan lacks global symmetry, but it is full of local symmetries at different levels of scale, such as the three sub-wholes at the first level and the nine sub-wholes at the second levels. These local symmetries are mainly created by numerous walls or thick boundaries as one of the 15 properties. Each of the local symmetries adapts its individual local context or need to make a better space. Overall, the plan has a good shape, because it comprises many good shapes at different levels of scale in an iterative or recursive manner. Some of the shapes echo each other in the overall configuration of the space. Eventually the large number of local symmetries and many of the other 15 properties recurring in the plan contribute a great deal to the life of the individual centers and the plan as a whole. The plan was likely to be designed unselfconsciously (Alexander 1964), but no one can deny the presence of the 15 fundamental properties. In this regard, the plan could be thought of as generated by iteratively applying the 15 properties, namely transformation properties, to the whole by differentiation and adaptation in a step-by-step fashion. The space continues to be differentiated with a wide range of scales from the smallest to the largest. Each step creates the context for the next one, and each wholeness develops, or more truly unfolds, from the previous one. This generation process underlies the structure-preserving or wholeness-extending transformations (Alexander 2002–2005).

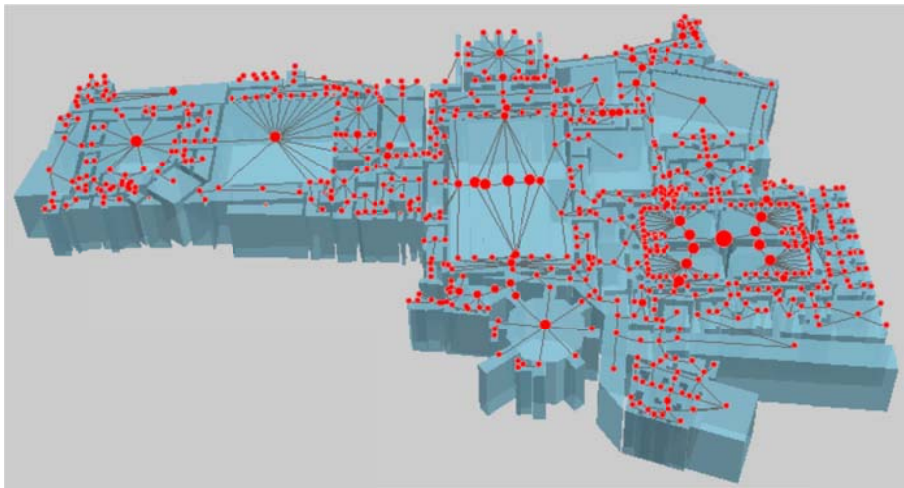

Figure 5: (Color online) The complex network of the 720 centers and 880 relationships of the Alhambra plan (Note: It is essentially the density of the centers, and their complex relationships that make the plan alive and beautiful, forming so called living structure or structural beauty. Every center is well adapted to its surrounding, and there are far more small centers than large ones, which arises from the continuous or iterative differentiation processes. The complex network as a whole is no less ordered than the rigid tree (Alexander 1965).)

**4.3 The Sierpinski carpet**
The Sierpinski carpet, although being one of many strictly defined fractals (Mandelbrot 1982), possesses a high degree of wholeness. The wholeness of the carpet or the coherence can also be assessed from two aspects: across all scales, and at one scale. Across all scales, the number of squares meets a power-law relationship with the levels of scale (Salingaros 1999), which should be extended to far more small things than large ones (see Section 5 for further discussion on spatial heterogeneity). At each scale, all pieces are the same size, which should be extended to more or less similar sizes (see Section 5 for further discussion of spatial dependence). To be more specific, there are three scales: 1/3, 1/9, and 1/27. Each of these scales respectively contains one, eight, and 64 squares, with a fractal dimension of $\log(8)/\log(1/3) = 1.89$ (Mandelbrot 1982). The 73 squares of the three different scales constitute a complex network as a whole (Panel b of Figure 6). With the complex network, there are two kinds of relationships: those among a same scale that are undirected, and those between two consecutive scales that are directed. The surrounding centers usually support the center ones, or the centers ones are enhanced by neighboring centers. In addition to the support relationships, there are also nested relationships among the 73 induced centers (Panel c of Figure 6). Because of these relationships, the central square has the highest degree of wholeness, which is accumulated from the eight middle squares and the 64 smallest squares. Eventually, the underlying scaling hierarchy of the Sierpinski carpet is shown in a tree structure (Panel d of Figure 6). It should be noted that the whole is not a tree, but a complex network.

Despite its high degree of wholeness, the Sierpinski carpet is not exactly the type of pattern we look for in practical designs. Instead, we seek the kind of pattern like the Alhambra plan. There are some major differences between the two patterns. The Sierpinski carpet is by creation, while the Alhambra is by step-by-step generation; the Sierpinski carpet has local and global symmetries, while the Alhambra lacks global symmetry, but with only local symmetries; all centers (or squares) of the Sierpinski carpet are the same size in one scale, while they are more or less similar in the Alhambra; and all shapes are squares in the Sierpinski carpet, while the Alhambra has different shapes. In addition, the Sierpinski carpet is too rigid in terms of the scaling ratio at precisely 1/3, and the increment of the number of squares across two consecutive scales is by exactly eight times. However, the two patterns are the same at the fundamental level in terms of wholeness or coherence at both local and global scales. The wholeness of the Alhambra can only be obtained through step-by-step creation or unfolding: differentiation and adaptation. In the course of design, a whole is continuously divided and differentiated to meet the scaling hierarchy, and newly added centers or scales are created to fit to their local contexts. Design or making is much more challenging than understanding, but the 15 properties (Alexander 2002–2005, Salingaros 2013, Salingaros 2015) provide guidance for structure-preserving transformations toward a whole with a higher degree of wholeness.

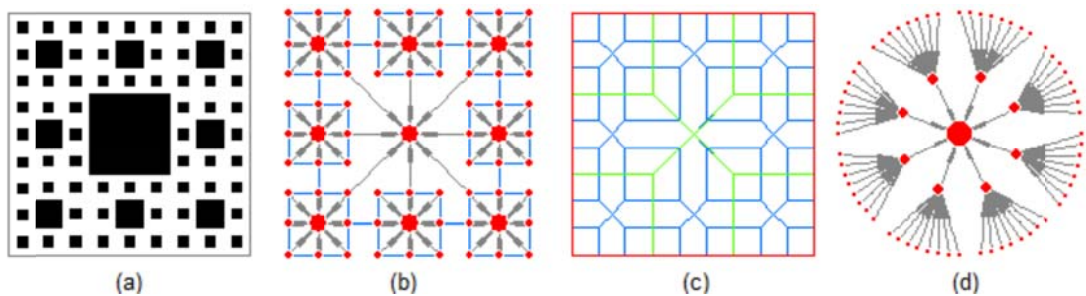

Figure 6: (Color online) A complex network perspective on the Sierpinski carpet
(Note: (a) the carpet with three scales, (b) its complex network of 73 pre-existing centers, in which the dot sizes represent degrees of in-links, (c) nested relationships among the 73 created centers, and (d) the scaling hierarchy of the complex network, in which the dot size represents the strength of the 73 pre-existing centers or their degrees of life.)

Through the case studies, the seemingly abstract concepts of wholeness and centers become more concrete and visible, helping reduce the mystery of wholeness. The wholeness comprises many recursively defined centers, and the centers are induced or created by the wholeness. For the sake of simplicity, we adopted these simple cases to illustrate the subtlety of the wholeness. The next section further discusses the implications of the complex network perspective to argue why the wholeness or living geometry is powerful and unique in terms of planning and repairing our environment or making the Earth's surface more whole or more beautiful.

## 5. Implications of the complex-network perspective and wholeness
The complex-network perspective enables us to see things in their wholeness, rather than as parts or fragments. Complex networks provide a powerful means to further study wholeness and understand the kind of problem a city is (Jacobs 1961). A city is essentially problems of organized complexity like those in biology. The notion that a city is more whole than another, or that a city has a high degree of wholeness, is the same as one city being more imaginable or legible than another (Lynch 1960, Jiang 2013b). Seen from the wholeness or the theory of centers, it is the city itself or the underlying living structure that determines the image of the city. The image of the city for individuals may vary from one to another, but there is a shared image of the city for all people. This shared image is the most interesting part for urban design theory, which has been persistently criticized for lacking a solid and robust scientific underpinning (Jacobs 1961, Marshall 2012). The complex network perspective or more truly the wholeness itself will inject scientific elements into urban planning and design.

The Sierpinski carpet is an image of the Earth's surface metaphorically. This is because there are far more small things than large ones across all scales – the spatial property of heterogeneity, and related things are more or less similar in terms of magnitude at each scale – the spatial property of dependence. Both heterogeneity and dependence are commonly referred to as spatial properties about geographic space or the Earth's surface (Anselin 1989, Goodchild 2004). The carpet is also an ideal metaphoric image toward which our built environment should be made. It implies that any space ought to be continuously differentiated to retain the scaling hierarchy across all scales ranging from the smallest to the largest, and any building or city ought to be adapted to its natural and built surroundings at each scale. The two processes of differentiation and adaptation enable us to create a whole with a high degree of wholeness. In this regard, the wholeness or the theory of centers would have enormous effects on geography, not only for better understanding geographic forms and processes, but also for planning and repairing geographic space or the Earth's surface (Mehaffy and Salingaros 2015, Mehaffy 2007). Built environments must adapt to nature, and new buildings must adapt to their surroundings.

The two spatial properties of heterogeneity and dependence constitute a true image of the Earth's surface at both global and local scales. Globally, spatial phenomena vary dramatically with far more small things than large ones across all scales, while locally they tend to be dependent or auto-correlated with more or less similar things nearby. These two properties are the source of the two kinds of harmony or coherence respectively across all scales and at every scale. For example, there are far more small cities than large ones across all scales globally (Zipf 1949), whereas nearby cities tend to be more or less similar in terms of the central place theory (Christaller 1933, 1966). However, the geography literature focuses too much on dependence, formulated as the first law of geography (Tobler 1970), but very little on heterogeneity. More critically, spatial heterogeneity is mainly defined for spatial regression with limited variation, governed by Gaussian thinking (Jiang 2015d). This understanding of spatial heterogeneity is flawed, given the fractal nature of geographic space or features. Spatial heterogeneity should be formulated as a scaling law because it is universal and global.

The wholeness or the theory of centers in general brings a new perspective to the science of complex networks. For example, the detection of communities of complex networks can benefit from the wholeness as a recursive structure. Instead of a flat hierarchy of community structures, a complex network contains numerous nested communities of different sizes, or far more small communities than large ones (Tatti and Gionis 2013, Jiang and Ma 2015). This insight about nested communities can be extended to classification and clustering. Current classification methods can be applied to mechanical assembly, so mechanical parts can be obtained. For a complex entity such as complex networks, the parts are often overlapping and nested inside each other. This resembles both wholeness and centers as a recursive structure. From a dynamic point of view, a complex network is self-evolved with differentiation and adaptation. A complex network has the power to evolve toward more coherence at both global and local scales, and is self-organized from the bottom up, rather than imposed from the top down. This insight about self-organization reinforces Alexander's piecemeal design approach through step-by-step unfolding or transformations for a living structure with a high degree of wholeness.

## 6. Conclusion
This paper develops a complex network perspective on the wholeness to make it more accessible to both designers and scientists. I discussed and compared related concepts such as whole and parts versus wholeness and centers to make them explicitly clear. A whole is a relatively coherent spatial set, while wholeness is a life-giving structure – *not something about the way they are seen, but something about the way they are* (Alexander 2003, p. 14). The wholeness is made of centers rather than arbitrarily identified parts. The centers are created or induced by the wholeness and made of other centers, rather than just those pre-existing in the whole. Given these differences, the mantra that *the whole is more than the sum of its parts* should be more correctly rephrased as t*he wholeness is more than the sum of its centers*. I demonstrated that the complex-network perspective enables us to see things in their wholeness. More importantly, I elaborated on design or making living structure through wholeness-extending transformation or unfolding in step-by-step differentiation and adaptation. Although understanding the nature of wholeness is essential, the ultimate goal of Alexander's concept of wholeness or living geometry is to allow us to make artifacts and built environments with the same order and beauty of nature itself.

The complex-network perspective enables us to develop new insights into planning and repairing geographic space. For example, differentiation and adaptation are two major processes for making living structures or geographic space in particular. This study showed that these processes underlie the two unique properties of geographic space: spatial heterogeneity and spatial dependence. These are respectively formulated as the scaling law and the first law of geography. Globally across all scales, there are far more small things than large ones. In contrast, locally at every scale, things tend to depend on each other with more or less similar sizes. These two spatial properties are the source of the two types of coherence respectively at global and local scales. In this connection, the theory of centers or living geometry would significantly contribute to understanding and making geographic space. Our future work points to this direction on how to rely on wholeness-extending transformations to create geographic features with a high degree of wholeness.

**Acknowledgement**
XXXXXXXXXX